\documentclass[12pt]{article}
\usepackage{amsbsy,amsmath}
\usepackage{amsfonts}
\usepackage{amssymb}
\usepackage{amscd}
\usepackage{bbm}
\usepackage{fancybox}
\usepackage{cite}
\usepackage{amsmath,amsfonts,amsbsy}
\usepackage{pstricks,pst-node}
\usepackage[small,bf,hang]{caption2}
\usepackage{graphicx}
\usepackage{epsfig}
\usepackage{psfrag}
\usepackage{comment}

\usepackage{float}

\psset{unit=1.3cm,linewidth=.5pt,radius=.2}  

\usepackage{multirow}                     
\usepackage{float}                          
\usepackage{lscape}                         
\usepackage{bm}

\newcommand{\bb}{\bar\beta}

\newcommand{\beq}{\begin{equation}}
\newcommand{\eeq}{\end{equation}}
\newcommand{\bi}{\begin{itemize}}
\newcommand{\ei}{\end{itemize}}
\newcommand{\bt}{\begin{tabular}}
\newcommand{\et}{\end{tabular}}
\newcommand{\bc}{\begin{center}}
\newcommand{\ec}{\end{center}}

\newcommand{\be}{\begin{equation}}
\newcommand{\ee}{\end{equation}}
\newcommand{\bea}{\begin{eqnarray}}
\newcommand{\eea}{\end{eqnarray}}
\newcommand{\ba}{\begin{array}}
\newcommand{\ea}{\end{array}}

\def\bbox{{\,\lower0.9pt\vbox{\hrule \hbox{\vrule height 0.2 cm
\hskip 0.2 cm \vrule height 0.2 cm}\hrule}\,}}
\newcommand{\dsl}{\pa \kern-0.5em /}

\font\mybb=msbm10 at 12pt
\def\bb#1{\hbox{\mybb#1}}

\def\bI {\bb{I}}

\def\tr{{\rm tr}}

\makeatletter \@addtoreset{equation}{section} \makeatother

\def\slashchar#1{\setbox0=\hbox{$#1$}           
   \dimen0=\wd0                                 
   \setbox1=\hbox{/} \dimen1=\wd1               
   \ifdim\dimen0>\dimen1                        
      \rlap{\hbox to \dimen0{\hfil/\hfil}}      
      #1                                        
   \else                                        
      \rlap{\hbox to \dimen1{\hfil$#1$\hfil}}   
      /                                         
   \fi}

\begin{document}

\begin{titlepage}

\begin{center}

\vskip 1.5cm

{\Large \bf On chiral bosons in 2D and 6D}

\vskip 1cm

{\bf Luca Mezincescu\,${}^1$ and  Paul K.~Townsend\,${}^2$} \\

\vskip 25pt

{\em $^1$  \hskip -.1truecm
\em Department of Physics, University of Miami, P.O. Box 248046, \\
Coral Gables, FL 33124, USA
 \vskip 5pt }

{email: {\tt mezincescu@miami.edu}} \\

\vskip .4truecm

{\em $^2$ \hskip -.1truecm
\em  Department of Applied Mathematics and Theoretical Physics,\\ Centre for Mathematical Sciences, University of Cambridge,\\
Wilberforce Road, Cambridge, CB3 0WA, U.K.\vskip 5pt }

{email: {\tt pkt10@cam.ac.uk}} \\

\end{center}

\vskip 0.5cm
\begin{center} {\bf ABSTRACT}\\[3ex]
\end{center}

In the Hamiltonian formulation of chiral $2k$-form electrodynamics, the $2k$-form potential on the $(4k+1)$-space is 
defined up to the addition of either (i) a closed $2k$-form or (ii) an exact $2k$-form, depending on the choice of chirality 
constraint. Case (i) is realized  by the Floreanini-Jackiw 2D chiral boson  (for $k=0$) and its Henneaux-Teitelboim 
generalisation to $k>0$.  For all $k$, but focusing on the 6D case, we present a simple Lorentz-invariant 
Hamiltonian model that realizes case (ii),  and we  derive it from Siegel's manifestly Lorentz invariant Lagrangian 
formulation.

\vfill

\end{titlepage}
\tableofcontents

\section{Introduction}

The term ``chiral boson''  is usually taken to mean a scalar field  in a two-dimensional (2D) locally-Minkowski spacetime
whose excitations travel (at light speed) only to the left (or to the right),  but it could equally well be taken to mean (for any non-negative integer $k$) a $2k$-form electrodynamics theory in a $(4k+2)$-dimensional locally-Minkowski spacetime  for which the $(2k+1)$-form field-strength is self-dual (or antiself-dual) \cite{Marcus:1982yu}. For $k=0$, and 2D Minkowski metric 
\be
ds^2_2 = - dt^2 + d\sigma^2 \, , 
\ee
the self-duality constraint on the one-form $d\phi$ for scalar field $\phi(t,\sigma)$ is
\be\label{CBeq}
\partial_-  \phi =0 \, , \qquad \partial_\pm := \frac{1}{\sqrt{2}} \left(\partial_t \pm \partial_\sigma\right)\, ,  
\ee
which implies that $\phi$ is a superposition of left-moving light-speed waves.

However, the equation \eqref{CBeq} is not the only Lorentz-invariant free-field equation for a 2D chiral boson; another possibility is the 
Floreanini-Jackiw equation:
\be\label{FJeq}
\partial_- \phi' =0\, , \qquad (\dot\phi, \phi') := (\partial_t\phi, \partial_\sigma\phi). 
\ee
This equation is not manifestly Lorentz invariant, because it is the space-derivative of \eqref{CBeq},  
but it {\it is} Lorentz invariant \cite{Floreanini:1987as}. It is also invariant under the additional restricted gauge transformation
\be\label{FJgt}
\phi(t,\sigma) \to \phi(t,\sigma) + \alpha(t)\, , 
\ee
where $\alpha$ is an arbitrary constant on the one-dimensional space but an arbitrary function of time.  In an application to the heterotic string, in which context $\phi(t,\sigma)$ represents the displacement of the string worldsheet in one of 16 `extra' space dimensions, the additional restricted gauge invariance implies that the string's centre of mass 
(and hence all particle-like oscillation modes) can move only in the nine `physical' space dimensions \cite{Townsend:2019koy}.

This distinction, based on gauge invariance, between the two 2D chiral boson equations is also a feature of the actions
that give rise to these equations by variation. The chiral boson equation \eqref{CBeq} is equivalent to the Euler-Lagrange (EL)
equations for the 2D `Siegel' action with Lagrangian density \cite{Siegel:1983es}
\be\label{Siegel}
\mathcal{L}_{\rm Siegel} =  \gamma\, \partial_-\phi \, \partial_+\phi   + \frac12 \lambda^{--} (\partial_-\phi)^2 \, , 
\ee
where $\lambda^{--}(t,\sigma)$ is a Lagrange multiplier field and $\gamma=1$, originally, but here we allow it to be an arbitrary constant. 
The  joint field equations of $\phi$ and $\lambda^{--}$ are equivalent to \eqref{CBeq} for {\it any} $\gamma$, including $\gamma=0$, because $(\partial_-\phi)^2=0$ implies $\partial_-\phi=0$.  Although the action depends on the Lagrange multiplier $\lambda^{--}$ in addition to the scalar field $\phi$, it is invariant under the 
following  ``Siegel symmetry'' gauge transformation with parameter $\alpha^-$  \cite{Siegel:1983es}:
\be\label{SSym}
\delta\phi= - \alpha^-\partial_-\phi  \, , \qquad 
\delta\lambda^{--} =  2\gamma\, \partial_+\alpha^-  + \lambda^{--}\partial_-\alpha^- -\alpha^-\partial_-\lambda^{--} \, . 
\ee
This is a ``trivial'' gauge transformation in the sense that the variation of the dynamical field $\phi$ is zero on-shell (i.e. on solutions of the
dynamical field-equation $\partial_-\phi=0$) but it allows a gauge to be chosen in which $\lambda^{--}$ is zero almost\footnote{A potential
legitimate gauge choice is $\lambda^{--} = c\delta(t-t_*)$ for some initial time $t_*$ and variable constant $c$, which imposes 
$\partial_-\phi=0$ as an initial condition preserved by the field equation $\square\phi=0$.} everywhere \cite{Siegel:1983es}. 

The Siegel action is {\it not} invariant under the (non-trivial)  restricted gauge transformation of \eqref{FJgt}, 
in contrast to the Floreanini-Jackiw action,  which has the Lagrangian density \cite{Floreanini:1987as}
\be\label{FJlag}
\mathcal{L}_{FJ} = (\partial_- \varphi)\varphi'\,  ,  
\ee 
with the FJ equation \eqref{FJeq} as its EL equation.   Although Lorentz invariance of $\mathcal{L}_{FJ}$ is not manifest, it can be made manifest in various ways; e.g. by the introduction of additional fields with additional gauge invariances\footnote{A well-known example is the PST method \cite{Pasti:1995us,Pasti:1996vs}. Another (string-inspired) example, in which Lorentz invariance becomes linearly realized as an `internal' symmetry,  can be found in \cite{Townsend:2019koy}.}. 

This distinction between 2D chiral boson theories explained above generalises
to chiral $2k$-form  electrodynamics in $(4k+2)$ dimensions. We shall focus here on the  $k=1$ 
case of  6D chiral 2-form electrodynamics, but we also explain how our results generalise to $k>1$. The 6D analog of the chiral boson equation \eqref{CBeq} is the manifestly 
Lorentz-invariant self-duality condition 
\be\label{CB6D}
0= \mathcal{F}^+ := \mathcal{F}  + \star \mathcal{F}\, , 
\ee
where $\mathcal{F}=d\mathcal{A}$ is the 3-form field-strength and $\star \mathcal{F}$ is its Hodge dual. In Minkowski coordinates 
$(t,\sigma^i; i=1,\dots,5)$, we may write $\mathcal{A}$ as
\be
\mathcal{A} =  \bb{A}_i\,  dt \wedge d\sigma^i + \frac12 A_{ij}\,  d\sigma^i\wedge d\sigma^j\, , 
\ee
so $\mathcal{A}_{0i}=\bb{A}_i$ and $\mathcal{A}_{ij} = A_{ij}$, and $\mathcal{F}$ is invariant under the gauge transformations 
\be\label{stang}
A_{ij} \to A_{ij} + 2\partial_{[i} \alpha_{j]} \, , \qquad \bb{A}_i  \to \bb{A}_i + \dot \alpha_i - \partial_i \alpha_0\, , 
\ee
where the parameters $(\alpha_0,\alpha_i)$ are the components of an arbitrary one-form on the 6D spacetime.
We may define the `electric' and `magnetic' components of $\mathcal{F}$ as follows:
\be\label{E&B}
\begin{aligned} 
E_{ij} &:= \mathcal{F}_{0ij} \equiv  \dot A_{ij} - 2\partial_{[i}\bb{A}_{j]} \, ,  \\
B^{ij} &:= \frac16\varepsilon^{ijklm} \mathcal{F}_{klm} \equiv  (\nabla \times A)^{ij} \, , 
\end{aligned} 
\ee
where we use a notation for which, for  any 5-space 2-form $C$, 
\be\label{5curl}
(\nabla \times C)^{ij} := \frac12 \varepsilon^{ijklm} \partial_k C_{lm} \, . 
\ee
The self-duality condition \eqref{CB6D} may now be written as  
\be\label{standardSD}
E = B \, , 
\ee
which has an  immediate $k>1$ generalization \cite{Henneaux:1988gg}.  This equation is Lorentz invariant because the 
infinitesimal Lorentz-boost transformations of $(E,B)$ are such that 
\be\label{LBEB}
\delta_\omega (E\pm B) = \pm \, \omega \times (E\pm B) \, ,
\ee
where $\omega$ is the constant 5-vector boost parameter, and the 5-space cross product of 
 $\omega$ with $(E-B)$ is defined, in analogy to the 5-space `curl' of  \eqref{5curl}. For later use, we detail here
the 5-space tensor algebra notation used in this paper. 
\begin{itemize}
\item  Cross product. For any 5-vector $w$ and 2-forms $(C,C^\prime)$, 
\be
(w \times C)^{ij} := \frac12 \varepsilon^{ijklm} w_k C_{lm} \, , \qquad (C\times C^\prime)^i := \frac14 \varepsilon^{ijklm} C_{jk} C^\prime_{lm} \, . 
\ee
Notice that $C\times C^\prime= C^\prime \times C$. 
\item Dot product and norm. \be\label{dotprod} 
C\cdot C^\prime := \frac12 C^{ij} C^\prime_{ij} \, , \qquad |C|^2 := C\cdot C\, . 
\ee
This is a special case of the standard inner-product on forms of arbitrary degree.

\item Triple scalar product. From $(w, C,C^\prime)$ we may construct a scalar in two potentially different ways, 
but the following identity implies their equivalence:
\be\label{5id}
(w \times C) \cdot C^\prime  \equiv w \cdot (C\times C^\prime) \, . 
\ee
The parentheses in the expressions on either side are optional because neither $w\times (C\cdot C')$ nor $w\cdot C$ is defined. 
\end{itemize}

A manifestly Lorentz invariant action that yields the self-duality equation $E=B$ was proposed by Siegel \cite{Siegel:1983es}; we shall consider 
it in detail later. All we need to know for the moment is that the equation $E=B$ is invariant under the gauge transformations \eqref{stang}. 
By taking the 5-space `curl' of this equation we arrive (since $\nabla \times E\equiv \dot B$)  at the Henneaux-Teitelboim (HT) equation \cite{Henneaux:1988gg}
\be\label{6DFJ}
\dot B - \nabla\times B =0\, . 
\ee
This is an equation for  the 2-form $A$ alone,  invariant under the gauge transformation 
\be
A_{ij} \to A_{ij} + \beta_{ij}\, , \qquad \partial_{[i}\beta_{jk]}=0\, , 
\ee
where $\beta_{ij}$ are the components of an arbitrary  {\it closed} 5-space 2-form $\beta$, which is also an arbitrary function of time. In the special 
case that $\beta={\rm d}\alpha$ (where ${\rm d}$ is the 5-space exterior derivative) we recover the gauge transformation of $A$ in \eqref{stang}, and 
in this sense the gauge invariance of \eqref{6DFJ} is `enlarged' relative to that of 
\eqref{standardSD}. This is also a feature of the HT action for which \eqref{6DFJ} is the EL equation; its 
Lagrangian density is   \cite{Henneaux:1988gg}
\be\label{HTaction}
\mathcal{L}_{HT} =  \dot A\cdot B - |B|^2\, . 
\ee
There are various ways to verify that $\mathcal{L}_{HT}$ is (despite appearances) Lorentz invariant  \cite{Henneaux:1988gg}. Here we simply observe that
it is the free-field case of the general Lorentz-invariant phase-space Lagrangian density of this form  \cite{Bandos:2020hgy}. 

The HT self-duality equation is the 6D analog of the 2D FJ chiral-boson equation. Alternatively, one may view the FJ equation as the $k=0$ 
case of the generalization of the 6D HT equation to arbitrary $k$. As confirmation of this claim, we observe that the parameter of the 
FJ gauge transformation \eqref{FJgt} is a closed 1-space 0-form (i.e., a constant on space but still an arbitrary function of time). The
non-existence of any {\it exact} 1-space 0-form accords with the fact that the `standard' 2D chiral boson equation ($\partial_-\phi=0$) is not invariant
under this transformation.

The main purpose of this paper is to explore this distinction between two `types' of chiral $2k$-form  electrodynamics 
in the context of their Hamiltonian formulations\footnote{Specifically, phase-space formulations in which the 
$2k$-form potential on the $(4k+1)$-space is one of the canonical variables; it is possible to deduce an alternative canonical formalism
directly from the gauge-invariant field equations but this leads ultimately to the  FJ/HT formulation \cite{Giannakis:1997hv}.}. 
For $k=0$ we will be following in the footsteps of other authors who have found the 
time-reparametrization Hamiltonian formulation of both the FJ and Siegel formulations of the 2D chiral boson, and noticed that the
 Hamiltonian chirality constraint is linear in the first case and quadratic in the second case. However, the significance of this fact has not previously been properly 
 appreciated, in our opinion; we discuss this in the following section.  
 
 We then move on to 6D chiral 2-form electrodynamics i.e. $k=1$, limiting the discussion to the free field case. As we review here, the HT version  can be  seen as a 
 modification of the non-chiral theory to include a  {\it linear} phase-space chirality constraint.  A new result of this paper is the Hamiltonian formulation of Siegel's 
 6D chiral 2-form theory \cite{Siegel:1983es}; we show that the phase-space action is gauge-equivalent to one with a simple  quadratic chirality constraint, which 
leads to a considerable  simplification of the ``trivial'' but  non-linear ``Siegel symmetry'' gauge invariances. 
However, our main conclusion is that the different implementation of the chirality constraint in the Hamiltonian formulations of the 6D Siegel and HT 
chiral 2-form theories  leads (as for $k=0$) to different classical theories, at least for periodic boundary conditions.

\section{2D chiral bosons and Hamiltonian constraints}\label{sec:2DCB}

Consider the following first-order 2D Lagrangian density with independent auxiliary field $\pi(t,x)$: 
\begin{equation}\label{tildeL}
\tilde{\mathcal{L}}_{(2)}  = \pi (\dot\phi - \phi') - \frac{\mu}{2} (\pi -\gamma\phi')^2\, .
\end{equation}
The subscript $(2)$ is to remind us that the chirality constraint imposed by the Lagrange multiplier $\mu$ is quadratic in the 
fields $(\phi',\pi)$. Elimination of $\pi$ by its algebraic field equation  takes us back to the 2D Siegel Lagrangian density of \eqref{Siegel}, with 
\be
\lambda^{--} = \mu^{-1} -\gamma \, . 
\ee
Although this appears to show that equivalence of \eqref{tildeL} to \eqref{Siegel} requires $\mu\ne0$, the EL equations for 
\eqref{tildeL} are jointly equivalent to 
\be
\partial_-\phi=0\, , \qquad \pi= \gamma\phi'\, , 
\ee
with $\mu$ undetermined, and hence equivalent in dynamical content to  \eqref{CBeq}. We may write \eqref{tildeL} in the form
\be\label{tildeL2}
\tilde{\mathcal{L}}_{(2)} = \pi\dot\phi - \mathcal{H} - \mu\Phi \, , 
\ee
where 
\be
{\mathcal H} = \pi\phi' \, , \qquad \Phi = \frac12 (\pi -\gamma\phi')^2 \, . 
\ee
This shows that $\tilde{\mathcal{L}}_{(2)}$ is a phase-space Lagrangian density for phase-space fields $(\pi,\phi)$ with 
Hamilton density $\mathcal{H}$ and phase-space chirality constraint $\Phi\approx 0$, imposed by a Lagrange multiplier $\mu$.

We have used Dirac's ``weak equality'' notation ($\approx$) because it serves to remind us that constraints may differ in their PB relations with functions on phase-space even when they are equivalent as equations; i.e. when they have the same solution space. As we shall see below, this  point is nicely illustrated by consideration of the alternative 
linear constraint 
\be
\chi \approx 0\, , \qquad \chi:= \pi -\gamma \phi' \, . 
\ee
The equation $\Phi=0$ is equivalent to the equation $\chi=0$ but replacing $\Phi$ by $\chi$ in \eqref{tildeL2} yields the 
alternative phase-space Lagrangian density 
\be\label{altL}
\tilde {\mathcal L}_{(1)} = \pi (\dot\phi - \phi') - \mu(\pi- \gamma\phi')\, .  
\ee
Notice that $(\pi,\mu)$ now form a {\it pair} of auxiliary fields because their joint field equations are
\be
\pi= \gamma\varphi'\, , \qquad \mu= \partial_- \varphi\, , 
\ee
which allows their consistent elimination from the action. The result is
\be\label{backtoFJ}
\tilde{\mathcal L}_{(1)} \to  \gamma (\partial_-\varphi)\varphi'\, , 
\ee
which is the FJ Lagrangian density {\eqref{FJlag} for $\gamma=1$. 

A version of this analysis was used by Faddeev and Jackiw to argue for equivalence of the 2D Siegel and Floreanini-Jackiw
chiral boson theories \cite{Faddeev:1988qp}.  Starting with $\tilde{\mathcal{L}}_{(2)}$, for $\gamma=1$,  they argued that the equivalence of the equation $\Phi=0$ 
to $\chi=0$ justifies  the substitution $\pi\to\phi'$ which, as we have seen, leads to  $\mathcal{L}_{FJ}$. The problem with this argument 
is that the equation of one variable ($\mu$) is being used to solve for another one ($\pi$), which is fine in the field equations but back-substitution into the 
action is not guaranteed to yield an equivalent action, and is therefore not generally legitimate.  If the substitution were legitimate in this case it would be so 
for any $\gamma$ including $\gamma=0$, but for $\gamma=0$ it leads to a zero action. The resolution of this difficulty is that the substitution is not 
legitimate, and for this reason we are not required to choose the same value for $\gamma$ in ${\mathcal L}_{(1)}$ and  $\tilde {\mathcal L}_{(2)}$. Indeed, 
$\gamma=0$ is  not allowed in $\tilde{\mathcal L}_{(1)}$, whereas it {\it is} allowed in $\tilde {\mathcal L}_{(2)}$. 

In principle, the substitution $\pi\to \gamma\phi'$ might also be illegitimate in $\tilde{\mathcal L}_{(1)}$ but in this context
 it yields the the same result as the (legitimate) simultaneous elimination of the auxiliary pair $(\pi,\mu)$. This exception to the general rule may be understood from a 
path-integral  perspective: given a path integral  over the fields $(\phi,\pi;\mu)$ with a measure determined by $\tilde{\mathcal L}_{(1)}$ we can first do the functional integral over $\mu$ to get the delta-functional $\delta[\chi]$ in the measure; the functional integral over $\pi$ is then trivial and the result is (provided $\gamma\ne0$) 
a path-integral over $\phi$ with a measure determined by the FJ action.  An argument along these lines was used by Bernstein and Sonnenschein but for
$\tilde{\mathcal{L}}_{(2)}$ \cite{Bernstein:1988zd}. In essence, they assumed that $\Phi$ can be replaced by $\chi$ for reasons similar to those of
Faddeev and Jackiw; they then showed that the resulting path integral is equivalent to one determined by the FJ action. As we have seen, this conclusion is correct 
given the premise, but the premise is false because $\delta[\Phi]$ is not generally equivalent to $\delta[\chi]$, just as $\delta (x^2)$ in the measure of an integral 
over real number $x$ is not generally equivalent to $\delta(x)$. 
 
We conclude this discussion  by summarising how the distinction between the 2D Siegel and Floreanini-Jackiw chiral boson theories
arises in the Hamiltonian formulation from the different effects of quadratic and linear chirality constraints that are equivalent as 
equations. 
\begin{itemize}

\item $\Phi\approx0$. The PB relation of $\Phi(\sigma)$ with any phase-space function, including $\Phi(\sigma')$, is
zero on the constraint surface. The constraints are therefore ``first-class'' and they generate gauge transformations which, however, are
``trivial'' because they are zero on the constraint surface. The Hamiltonian field equations are equivalent to the standard chiral boson 
equation $\partial_-\phi=0$ for any value of the constant $\gamma$. 

\item $\chi \approx0$. For $\gamma\ne0$  the Hamiltonian field equations are equivalent to the FJ chiral boson equation. 
The infinite set of  Fourier modes of $\chi$ is a basis for mixed first and second class constraints. 
The non-zero modes come in pairs and they are a basis for a set of  second-class constraints, but the lone zero mode is first class 
and it generates the restricted gauge invariance of the FJ theory \cite{Townsend:2019koy}. 

\end{itemize}

Here we must stress that the above discussion concerns the classical field theory. In the quantum theory, 
the quadratic Siegel constraints are no longer first-class because of a quantum anomaly, so modifications are 
required \cite{Imbimbo:1987yt}, and complications associated with the second-class constraints of the FJ formulation 
have led to other modifications (e.g. \cite{McClain:1990sx,Devecchi:1996cp}).  However, none of these modifications negate the fact that the 
physical phase spaces of the Siegel and FJ formulations differ for periodic boundary conditions because of the additional 
restricted gauge invariance of the latter, and we may expect this difference to be important to the quantum theory. This expectation is 
confirmed by a Lagrangian-based derivation  of the partition function for a chiral boson on a 2-torus \cite{Chen:2013gca}.

\section{6D chiral 2-form electrodynamics}

For Minkowski coordinates $x^\mu= (t, \sigma^i)$, the manifestly Lorentz invariant free-field Lagrangian density for the {\it non-chiral} 6D 2-form electrodynamics is
\be\label{6DLagCov}
\mathcal{L} = -\frac{1}{12} \mathcal{F}^{\mu\nu\rho} \mathcal{F}_{\mu\nu\rho}  \qquad \left(\mathcal{F}_{\mu\nu\rho}= 3\partial_{[\mu} \mathcal{A}_{\nu\rho]}\right), 
\ee
where a Minkowski metric of `mostly plus' signature is used to raise Lorentz-vector indices. Using the definitions \eqref{E&B} of electric and magnetic 
components of $\mathcal{F}$, we find the alternative expression 
\be\label{6DLag}
\mathcal{L} =  \frac12(|E|^2 -|B|^2)\, .  
\ee
Only the invariance under 5-space rotations is now obvious but we still have a linearly-realized  invariance under the Lorentz boost transformations of 
\eqref{LBEB}, so the Lorentz invariance is still ``manifest'' in this sense. 

The EL equations obtained from \eqref{6DLagCov} or \eqref{6DLag} propagate a total of 6 modes in the $(0,3) \oplus (3,0)$ representation of the 6D lightlike
``little group'' $SO(4)$.  In contrast, the chiral 2-form equation $E=B$ (which, we recall, is the self-duality condition on the spacetime 3-form $\mathcal{F}$)
 propagates only three modes. The attempt to construct an  action for this equation by means of a Lagrange multiplier that imposes $E=B$  fails because the resulting action then includes 3 additional modes propagated by the Lagrange multiplier; this is because 
 $E=B$ is a dynamical equation, not a constraint.  This difficulty is most simply solved,  at the cost of a loss of {\it manifest} Lorentz invariance\footnote{It can also be 
 solved while maintaining manifest Lorentz invariance, as for 2D; however,  as pointed out in \cite{Bandos:2020hgy}, there is always some non-manifest symmetry or gauge invariance.},  by passing to the Hamiltonian formulation because a chirality condition may then be imposed as  a (non-dynamical) phase-space constraint. However, the Hamiltonian route to an off-shell chiral 2-form electrodynamics may be implemented in essentially two distinct ways: we may choose  the chirality constraint to be linear in the natural phase-space variables, or we may choose it to be quadratic.  These two options do not preclude the possibility of  more complicated 
non-linear chirality constraints but they are  the simplest choices that lead to the two possible 6D free-field chiral 2-form equations, 
distinguished by their distinct set of gauge invariances, in close analogy to our discussion of the 2D chiral boson.   

We shall begin with a review of  the Hamiltonian formulation  of non-chiral 2-form electrodynamics and how  the HT chiral 2-form theory is obtained 
from it by imposing a linear chirality constraint.  We then investigate the consequences of replacing this linear constraint by a simple quadratic 
one that preserves a non-linearly realized Lorentz invariance. We then derive this simple `quadratic' Hamiltonian formulation from Siegel's manifestly
Lorentz invariant Lagrangian formulation. This result exposes the close analogy of the  6D Siegel and HT formulations of 6D chiral 2-form electrodynamics
to the 2D Siegel and FJ formulations of the 2D chiral boson, in addition to greatly simplifying the ``Siegel symmetry'' gauge invariances of the Siegel formulation.

\subsection{Hamiltonian formulation: non-chiral case}

The Hamiltonian density for the {\it non-chiral} theory  is the Legendre transform of  the Lagrangian density of \eqref{6DLag} 
with respect to $E$: 
\be
\mathcal{H}(D,B) = \sup_E \left\{ D\cdot E - \mathcal{L}(E,B)\right\} = \frac12 \left(|D|^2+|B|^2\right) \, , 
\ee
where the antisymmetric tensor field $D^{ij}$ is the momentum variable canonically conjugate to $A_{ij}$. 
The Hamiltonian field equations are the Euler-Lagrange equations for the phase-space Lagrangian density 
\be\label{nonch} 
\tilde{\mathcal L} = E\cdot D - {\cal H}(D,B) \qquad ({\rm non-chiral}) \, . 
\ee
In this free-field case, the field equation for $D$ is $D=E$, and we then 
recover \eqref{6DLag} by (consistent) substitution.  Notice that 
\be\label{notice}
E\cdot D = \dot A \cdot D - \bb{A}_j \bb{G}^j - \partial_i( \bb{A}_jD^{ij}) \, , \qquad \bb{G}^i= \partial_jD^{ij}\, , 
\ee
which shows that $\bb{A}$ is a 5-vector Lagrange multipler for a (first class) 5-vector phase-space constraint; the 
5-vector constraint function $\bb{G}$ generates  the gauge transformation of the 5-space 2-form potential $A$. To verify this, it is useful to consider
a functional basis parametrised by smooth 1-forms $\alpha(\boldsymbol{\sigma})$; i.e.
\be\label{gausslaw}
G[\alpha] = \int\! d^5\!\sigma\, \alpha_i \bb{G}^i \, . 
\ee
Assuming that the 1-forms $\alpha$ have compact support (which allows us to ignore surface terms when integrating by parts)
we may use the canonical PB relations 
\be\label{canPB}
\left\{A_{ij}(\boldsymbol{\sigma}), D^{kl} (\boldsymbol{\tau}) \right\}_{PB} = 2\delta_{[i}^k \delta_{j]}^l\,\delta(\boldsymbol{\sigma}- \boldsymbol{\tau})\, , 
\ee
to show that $G[\alpha]$ has a zero PB with the Hamiltonian (because it is gauge-invariant) and that 
$\left\{G[\alpha], G[\alpha'] \right\}_{PB} = 0$. The set of functionals $\{G\}$ is therefore trivially first-class, and the infinitesimal gauge 
transformation generated by $G[\alpha]$ is
\be\label{gt1}
\delta_\alpha A_{ij} = \left\{A_{ij}, G[\alpha]\right\}_{PB} =  2\partial_{[i} \alpha_{j]}\, , 
\ee
which is the usual gauge transformation in which the 5-space 2-form potential $A$ is shifted by an exact 2-form $d\alpha$. 

As an aside on conventions, we remark that $2\delta_{[i}^k \delta_{j]}^l$ is the identity matrix acting on the 10-dimensional space of 
antisymmetric $5\times 5$ matrices since
\be
\frac12 \left[2\delta_{[i}^k \delta_{j]}^l\right] \left[ 2\delta_{[k}^p \delta_{l]}^q\right] = 2\delta_{[i}^p \delta_{j]}^q\, , 
\ee
where the factor of $\tfrac12$ is needed to avoid overcounting; this is the same factor of $\tfrac12$ that appears in  \eqref{dotprod}.

\subsection{Linear chirality constraint: the HT case}

A chiral version of the free-field 2-form electrodynamics may be found by using a Lagrange multiplier field to impose the linear chirality 
constraint \cite{Henneaux:1988gg}
\be
\chi_{ij} := (D-B)_{ij} \approx 0\, . 
\ee
This yields the following phase-space Lagrangian density:
\be\label{HT+}
\tilde{\mathcal L}_{(1)} = \dot A\cdot D - \frac12 \left(|D|^2 + |B|^2\right) - \mu\cdot \chi\, , 
\ee
where the subscript ``$(1)$'' serves to remind us that the chirality constraint imposed by the 5-space 2-form Lagrange multiplier $\mu$ is 
linear in the phase-space fields $(E,D)$.   We have omitted the $\bb{A}\cdot \bb{G}$ term because it can be removed (ignoring total derivatives) 
by a redefinition of the Lagrange multiplier $\mu$;  the `Gauss law' constraint $\bb{G}\approx 0$ is thus subsumed into the chirality constraint.

The 2-form fields $(D,\mu)$ form an auxiliary pair that may be eliminated by their joint algebraic field equations. 
This yields the Henneaux-Teitelboim chiral 2-form Lagrangian density of \eqref{HTaction}. Direct substitution $D\to B$ yields the same result, as in the 2D case for a linear chirality constraint,  and this can again be explained by path-integral considerations. The Lagrangian density $\tilde{\mathcal L}_{(1)}$ is therefore 
equivalent to $\mathcal{L}_{HT}$ of  \eqref{HTaction}, but has the advantage that its gauge invariances are now associated to first-class 
constraints, as we now explain. 

As for the non-chiral theory, it is convenient to choose a functional basis for the constraints, which are 
now parametrised by 5-space 2-forms $\beta_{ij}$:
\be
\chi[\beta] = \int\! d^5\sigma\, \beta \cdot\chi\, . 
\ee
Using the canonical Poisson bracket relations 
\be
\left\{ A_{ij}(\boldsymbol{\sigma}), D^{kl}(\boldsymbol{\sigma'}) \right\}_{PB} = 
2\delta_{[i}^k\delta_{j]}^l\,  \delta(\boldsymbol{\sigma} - \boldsymbol{\sigma'})\, , 
\ee
which implies\footnote{The substitution $D\to B$ in this PB relation does not give the correct result for
the PB relations involving only components of $B$; these are all zero in the current context, but the 
corresponding Dirac brackets are (necessarily) the same as the Poisson brackets 
deduced directly from  \eqref{HTaction}.}
\be\label{PBDB}
\left\{ B^{ij}(\boldsymbol{\sigma}), D^{kl}(\boldsymbol{\sigma'}) \right\}_{PB} = \varepsilon^{ijklm} \partial_m 
 \delta(\boldsymbol{\sigma} - \boldsymbol{\sigma'})\, , 
\ee
we find that \cite{Townsend:2019ils}
\be
\left\{\chi[\beta], \chi[\beta'] \right\}_{PB} = 2\!\int\! \beta \wedge {\rm d}\beta' \, ,  
\ee
which shows that we have a set of mixed first-class and second-class constraints. The first-class subset 
is parametrized by {\it closed} 2-forms $\beta$, and these generate the infinitesimal gauge transformations of $A$:
\be\label{gt2}
\delta_\beta A_{ij} = \left\{A_{ij}, \chi[\beta]\right\}_{PB} = \beta_{ij} \, , \qquad \partial_{[k}\beta_{ij]} =0\, . 
\ee
For $\beta={\rm d}\alpha$ we recover the gauge transformation \eqref{gt1} but now we have a larger set of gauge transformations
because the closed 2-forms $\beta$ include the harmonic 5-space 2-forms, as has been previously emphasized in \cite{Bekaert:1998yp}, 
and in \cite{Bandos:2014bva} in the context of a PST-covariantization of the HT action. 

Whether there are any harmonic 2-forms depends on the boundary conditions; these have been left unspecified but there certainly 
are harmonic 2-forms if we impose periodic boundary  conditions in at least two independent directions. For example, we could take 
the 5-space to be a flat 5-torus $T^5$ or $\bb{R}\times T^4$,  in which case there is a consistent dimensional-reduction/truncation to the 
FJ chiral boson on $S^1$ or $\bb{R}$, 
respectively  \cite{Bandos:2020hgy}.

\subsection{Interlude: a quadratic chirality constraint}\label{ssec:3quad}

Possibly the simplest action with the self-duality condition $E=B$ as its EL equations is the action with Lagrangian density
\be\label{Lquad}
\mathcal{L} = \frac{\gamma}{2} \left(|E|^2-|B|^2\right) + \frac{\lambda}{2}\,  |E-B|^2\, , 
\ee
where $\gamma$ is an arbitrary constant (which we could set to zero) and $\lambda$ is a (rotation scalar) Lagrange multiplier. 
The field equation found from variation of $A$ is implied by the equation found from variation of  $\lambda$, which is $E=B$ (because
$|E-B|=0$ iff $E-B=0$).  The action is also Lorentz invariant, but this is not manifest because it is realized non-linearly on the Lagrange multiplier
$\lambda$. To verify this, we  observe that the Lorentz-boost transformations of \eqref{LBEB} yield 
\be\label{LBlam}
\delta_\omega |E-B|^2 = -2\omega\cdot (E-B)\times (E-B)\, , 
\ee
from which it follows that  the Lagrange multiplier term of \eqref{Lquad} is Lorentz invariant if the  infinitesimal Lorentz-boost transformation of 
$\lambda$ is taken to be
\be\label{lamboost}
\delta_\omega \lambda = 2\lambda\,  \omega\cdot N\times N\,  ,  
\ee
where
\be
N= (E-B)/|E-B|\, . 
\ee
Although $N$ is ambiguous when $E=B$, the product $\delta_\omega\lambda |E-B|^2$ is unambiguously zero when $E=B$. It also follows from \eqref{LBEB} that 
\be\label{LBN}
\delta_\omega N= - \omega \times N - (\omega\cdot N\times N) N\, , 
\ee
and this may be used to verify that the commutator of two Lorentz boosts acting on $\lambda$ with 5-vector parameters $\omega$ and $\omega'$ is zero, as it should be since $\lambda$ is a rotation scalar. 

Our motivation for analysing the model defined by \eqref{Lquad} is that it is a useful preliminary to an analysis 
of Siegel's manifestly Lorentz invariant Lagrangian formulation because it is possible to bring the Siegel Lagrangian density to the form 
\eqref{Lquad} by a partial fixing of the ``Siegel'' gauge invariances; this will be demonstrated in the following subsection. The infinitesimal transformations 
of the residual  Siegel gauge invariance are 
\be\label{trivialgt}
\begin{aligned} 
\delta_\xi A &= -\xi (E-B) \, , \\
\delta_\xi\lambda &= \gamma(\dot\xi + \nabla\xi\cdot N\times N) + \lambda\overleftrightarrow{\partial_t}\xi - 
\lambda \overleftrightarrow{\nabla} \xi \cdot N\times N \, , 
\end{aligned} 
\ee
where we use the notation
\be
\lambda \overleftrightarrow{\rm d} \xi \equiv \lambda {\rm d}\xi - \xi {\rm d}\lambda\, . 
\ee 
Notice that the variation of $A$ is zero ``on-shell'', making this a ``trivial'' gauge invariance.  In addition both $\lambda$ and the parameter $\xi$ 
are arbitrary functions of both time and space, so we may choose a gauge in which $\lambda$ is zero almost everywhere.  For our purposes, 
gauge invariances with these two properties are  ``Siegel'' gauge invariances.  

We give here some details of the proof that the transformations 
\eqref{trivialgt} leave invariant the Lagrangian density of \eqref{Lquad} if we ignore total derivative terms in its variation. This will also serve to 
illustrate the utility of the 5-vector algebra notation summarized in the Introduction: 
\begin{itemize}

\item The variation $\delta_\xi A= -\xi(E-B)$ of \eqref{trivialgt} implies that 
\be
\delta_\xi E= -\partial_t\left[\xi(E-B)\right] \, , \qquad \delta_\xi B= -\nabla\times \left[\xi(E-B)\right]\, , 
\ee
and hence 
\be 
\begin{aligned}
\delta_\xi\left\{ \frac12\left(|E|^2-|B|^2\right) \right\} &= -E\cdot \partial_t \left[\xi(E-B)\right] + B\cdot \nabla\times \left[\xi(E-B)\right]\\
&= \xi\left[ \dot E - \nabla\times B\right] \cdot (E-B)  + \  {\rm total\ derivative} \, . 
\end{aligned}
\ee
Omitting the total derivative we then have
\be
\begin{aligned}
\delta_\xi\left\{ \frac12\left(|E|^2-|B|^2\right) \right\} &= \xi(E-B)\cdot \left[\partial_t(E-B) + (\dot B- \nabla\times B)\right]\\
&= \xi (E-B) \cdot \left(\partial_t + \nabla \times\right) (E-B) \\
&= \frac{\xi}{2} \left\{\partial_t |E-B|^2 + \nabla\cdot  \left[ (E-B)\times (E-B)\right] \right\}\, . 
\end{aligned}
\ee
Integrating by parts and again omitting total derivatives, we arrive at 
\be
\delta_\xi\left\{ \frac12\left(|E|^2-|B|^2\right) \right\} = -\left(\dot\xi + \nabla\xi\cdot N\times N\right) \left[ \frac12|E-B|^2\right] \, . 
\ee
Next we compute
\be
\begin{aligned}
\delta_\xi \left\{\frac12 |E-B|^2\right\} &= -(E-B) \cdot (\partial_t - \nabla\times)\left[\xi (E-B)\right]\\
&= -\left(\dot\xi - \nabla\xi \cdot N\times N\right)|E-B|^2 \\
& \quad - \frac{\xi}{2} \left\{ \partial_t|E-B|^2- \nabla \cdot \left[(E-B)\times (E-B)\right] \right\} \, , 
\end{aligned}
\ee
and hence, omitting total derivatives, 
\be
\begin{aligned}
\delta_\xi \left\{\frac{\lambda}{2} |E-B|^2\right\} &= \left[ \delta_\xi\lambda - 2\lambda(\dot\xi - \nabla\xi \cdot N\times N)\right] \left[ \frac12|E-B|^2\right] \\
&\quad + \left[\partial_t(\lambda\xi) - \nabla(\lambda\xi) \cdot N\times N\right]   \left[ \frac12|E-B|^2\right] \\
&= \left[ \delta_\xi\lambda - \lambda\overleftrightarrow{\partial_t}\xi + \lambda\overleftrightarrow{\nabla}\xi \cdot N\times N\right]  \left[ \frac12|E-B|^2\right]\, . 
\end{aligned}
\ee
Putting these results together we find that the variation of $\mathcal{L}$ is a total derivative provided that $\delta_\xi\lambda$ is given by the expression in \eqref{trivialgt}.

\end{itemize}

We now turn to the Hamiltonian formulation. Consider the following first order Lagrangian density 
\be\label{foquad}
\tilde{\mathcal{L}}_{(2)} = D\cdot (E - B) - \frac{\mu}{2} |D-\gamma B|^2\, , 
\ee
where $D$ is an independent auxiliary 5-space 2-form field, and 
\be\label{mutol}
\mu = \frac{1}{\lambda+\gamma}\, . 
\ee
The subscript ``$(2)$'' serves to remind us that the (rotation scalar) Lagrange  multiplier $\mu$  imposes  a {\it quadratic} chirality 
constraint. Elimination of $D$,  by means of its field equation
\be\label{Dfe}
D-\gamma B = \mu^{-1} (E-B) \, , 
\ee
yields the the second-order Lagrangian density of \eqref{Lquad}, so we now have an equivalent first-order version of it. 
Moreover, Lorentz invariance is preserved in the passage from the second order ${\mathcal L}_{(2)}$ to the 
first-order $\tilde{\mathcal L}_{(2)}$; the infinitesimal Lorentz-boost  transformation of $D$ is 
\be\label{Dom}
\delta_\omega D  =  (1-\gamma\mu)\,  \omega \times (D-\gamma B) + \gamma\,  \omega \times B\, , 
\ee
and the Lorentz-boost transformation of $\mu$ is
\be\label{muom}
\delta_\omega\mu = -2\mu(1-\gamma\mu)\,  \omega\cdot \tilde N\times \tilde N \, , \qquad \tilde N= (D-\gamma B)/|D-\gamma B|\, . 
\ee
On elimination of $D$  this transformation of $\mu$ implies that of $\lambda$ in \eqref{LBlam} since the $D$ field equation implies 
$\tilde N=N$.  In contrast, the variation \eqref{Dom} of $D$ as an independent field does {\it not} agree with its variation as the function 
of $(E,B,\mu)$ implied by \eqref{Dfe} unless we also use the dynamical equation $E=B$, but this fact is perfectly compatible
with the off-shell Lorentz-boost invariance of the Lagrangian density $\mathcal{L}_{(2)}$ obtained from $\tilde{\mathcal L}_{(1)}$
by substitution for $D$; it is just an indication that Lorentz-boost invariance
of the field equations of $\tilde{\mathcal L}_{(1)}$ is not achieved by a separate invariance of its algebraic and dynamical equations. 

We may rewrite \eqref{foquad} as 
\be\label{foquad2}
\tilde{\mathcal{L}}_{(2)} = E\cdot D  - \mathcal{H}   - \mu\, \Phi(D,B) \, ,  
\ee
where 
\be
\mathcal{H} = D\cdot B\, , \qquad \Phi := \frac12|D-\gamma B|^2 \, . 
\ee
We recognise this as a phase-space Lagrangian density with Hamiltonian density $\mathcal{H}$ and phase-space constraint 
$\Phi\approx0$. The surface in phase-space defined by this constraint is exactly the same as it was for the linear constraint;  i.e. 
$D=\gamma B$, and hence\footnote{Recall that $\approx$ indicates ``weak equality'' in Dirac's sense.}
\be
\tilde{\mathcal{H}}  \approx \left(\frac{\gamma}{\gamma+1}\right) \left(|D|^2 + |B|^2\right)\, . 
\ee
For $\gamma=1$ this is the Hamiltonian density of $\tilde{\mathcal{L}}_{(1)}$. 

We should not forget that $E\cdot D$  includes an $\bb{A}\cdot \bb{G}$ term, which can {\it not} now be removed by a redefinition of the 
Lagrange multiplier for the chirality constraint. We therefore have two (sets of) constraints, one with a basis of functionals $G[\alpha]$, for 
5-space 1-form parameters $\alpha= \alpha_i d\sigma^i$, and another with a basis of functionals
\be
\Phi[\beta] = \int\! d^5\sigma \, \beta(\boldsymbol{\sigma}) \Phi(\boldsymbol{\sigma})  \, , 
\ee
where the parameter  $\beta$ is an inverse-scalar-density; as before we assume that the parameters are smooth and have finite support
but are otherwise arbitrary. As in the non-chiral theory, $G[\alpha]$ generates the standard gauge transformation 
$A\to A + d\alpha$  of the 5-space 2-form potential $A$. As  $\Phi[\beta]$ is invariant under this gauge transformation, its Poisson bracket with 
$G[\alpha]$  is zero.

We still need the PB relations of the functionals $\Phi[\beta]$.  Using \eqref{PBDB} we find that 
\be
\left\{\Phi[\beta^\prime], \Phi[\beta] \right\}_{PB} = 
2\gamma\,  \Phi\left[ (\beta\overleftrightarrow{\nabla}\beta^\prime \cdot \tilde N\times \tilde N\right]\, , 
\ee
which shows that the functionals $\Phi[\beta]$ span a first-class set of constraints, and therefore generate gauge transformations of the canonical variables. 
To verify this claim we compute
\be
\begin{aligned}\label{betacan}
\delta_\beta D &\equiv& \left\{ D, \Phi[\beta]\right\}_{PB} &=  -\gamma\nabla \times \left[\beta(D-\gamma B)\right]  \\
\delta_\beta A &\equiv& \left\{ A, \Phi[\beta]\right\}_{PB} & = \beta (D-\gamma B)\, , 
\end{aligned}
\ee
from which we deduce 
\be
\delta_\beta E = \partial_t \left[\beta(D-\gamma B) \right] \, , \qquad 
\delta_\beta B  =  \nabla \times \left[\beta(D-\gamma B)\right]   \, . 
\ee
With these variations in hand, we find that $\delta_\beta \tilde{\mathcal{L}}_{(2)}=0$ (omitting total derivatives) provided that\footnote{Again, this variation is not defined when $D=\gamma B$ but the product $(\delta_\beta\mu)\Phi$ is unambiguous.}
\be
\delta_\beta\mu = \dot\beta  - \nabla\beta \cdot  \tilde N\times \tilde N \, . 
\ee
However, the $\beta$-gauge transformations of the canonical variables are weakly zero and hence do not reduce the dimension of the physical phase space; in other words, 
$\Phi[\beta]$ generates ``trivial'' gauge invariances that have no effect on the field equations. This trivial gauge invariance is just
the Hamiltonian version of the  invariance of the Lagrangian $\mathcal{L}$ under the gauge transformations \eqref{trivialgt} (the parameters are 
related by $\beta=-\mu\xi$).  The only non-trivial gauge invariance is the one generated by $G[\alpha]$, as in the non-chiral theory.

\subsection{Siegel's chiral 2-form electrodynamics}\label{sec:4}

The stress-energy tensor of the non-chiral 2-form electrodynamics theory may be written as the sum
\be
T_{\mu\nu} = T^-_{\mu\nu} + T^+_{\mu\nu} \, , \qquad 
T^\pm _{\mu\nu} :=\frac18 \mathcal{F}^{\pm}_{\mu \lambda\rho} \mathcal{F}^{\pm}_\nu{}^{\lambda\rho} \, .  
\ee
We get a chiral theory by setting to zero one of the two terms in the sum. Implementing this by a Lagrange multiplier,
we arrive at the manifestly Lorentz invariant Lagrangian density
\begin{equation}\label{6DS}
\mathcal{L} = -\frac{\gamma}{12} \mathcal{F}^{\mu\nu\rho} \mathcal{F}_{\mu\nu\rho} + 
 \lambda^{\mu\nu}T^+_{\mu\nu}  \, , 
\end{equation}
where $\gamma$ is a constant and $\lambda^{\mu\nu}$ a symmetric tensor Lagrange multiplier, which we may assume to be traceless 
(i.e. $\eta_{\mu\nu}\lambda^{\mu\nu}=0$, where $\eta$ is the Minkowski metric).  For $\gamma=1$ this is the Lagrangian density proposed by 
Siegel \cite{Siegel:1983es} (after a rescaling of the Lagrange multiplier) but the field equations are equivalent to the free-field 
chiral 2-form equation $\mathcal{F}^+=0$ for any value of $\gamma$, including $\gamma=0$.  As Siegel showed, there are gauge invariances
that are ``trivial'' but which act non-trivially on the traceless-tensor Lagrange multiplier field, such that it may almost be gauged away; this is the 6D analog of the 
2D ``Siegel symmetry'' with transformations \eqref{SSym}. 

To pass to the Hamiltonian formulation, we first  make a time/space split in order to rewrite \eqref{6DS} in terms of the electric and magnetic components of 
$\mathcal{F}$.  The corresponding components of $\mathcal{F}^+$ are 
\be\label{Fpluscomp}
\mathcal{F}^+_{0ij} = (M_-)_{ij}  \, , \qquad 
 \mathcal{F}^+_{ijk} = - \frac12  \varepsilon_{ijklm} (M_+)^{lm} \, ,  
\ee
where we use the notation 
\be
M_\pm = E\pm B\, . 
\ee
We now find that the Lagrangian density of \eqref{6DS}  is
\be
\mathcal{L}  =  \frac{\gamma}{2}  M_+\cdot M_-  + \frac12\left\{\lambda^{00}  |M_-|^2
 - \upsilon\cdot M_- \times M_- + \, \lambda^{ij} \left[M_-^2\right]_{ij}\right\} \, , 
\ee
where we use the notation 
\be
u^i = \lambda^{0i} \, , 
\ee
and $M_\pm^2$ is the matrix square of $M_\pm$ (so that $\tr M_\pm^2 = -2|M_\pm|^2$).  
As $\lambda^{00} = \lambda^k{}_k$, we have 
\be
\lambda^{ij} = \tilde\lambda^{ij} + \frac15 \delta^{ij} \lambda^{00} \, , \qquad \tilde\lambda^k{}_k \equiv 0\, , 
\ee
and we may rewrite the Lagrangian density as 
\be\label{6Dus+} 
\mathcal{L}  =  \frac{\gamma}{2}  M_+\cdot M_-  + \frac12\left\{\frac35 \lambda^{00}  |M_-|^2
 - \upsilon\cdot M_- \times M_- + \, \tilde \lambda^{ij} \left[M_-^2\right]_{ij}\right\} \, . 
\ee
In this form we see clearly that the original Lagrange multiplier, which transforms as an irreducible ${\bf 20}$ of the 6D Lorentz group, decomposes into the
${\bf 1}\oplus {\bf 5} \oplus {\bf 14}$ of the SO(5) rotation group.  The field equations are still equivalent to the standard 6D self-duality relation, 
which now takes the form $M_-=0$. 

In the current notation, the infinitesimal Lorentz-boost transformations of \eqref{LBEB} are
\be\label{LBMpm}
\delta_\omega M_\pm  =  \pm\,  \omega \times M_\pm \, . 
\ee
The infinitesimal Lorentz-boost transformations of the Lagrange multiplier fields are
\be\label{LBlms}
\delta_\omega \lambda^{00} = -2\omega\cdot \upsilon\, , \qquad \delta\upsilon^i = -2\, \omega^{(i} \lambda^{j)}{}_j\, , \qquad 
\delta_\omega\lambda^{ij}  = -2\, \omega^{(i} \upsilon^{j)}  \, . 
\ee
Notice that these are {\it linear} transformations; in this respect Lorentz invariance is still ``manifest''. 
To verify Lorentz boost invariance, it is convenient to use the identities
\bea\label{triplex}
M_\pm^{i[j} M_\pm^{kl]}  &\equiv&  \frac16 \varepsilon^{ijklm} (M_\pm\times M_\pm)_m\, , \nonumber \\
(\omega \times M_\pm)\times M_\pm  &\equiv & (M_\pm^2 + |M_\pm|^2 \bI_5)\omega \, , 
\eea
to verify that 
 \begin{eqnarray}\label{formulae}
 \delta_\omega (M_\pm^2)_{ij}  &=& \pm  \left[\omega_{(i} (M_-\times M_-)_{j)} -  (\omega\cdot M_-\times M_-)\delta_{ij} \right]\, , \nonumber \\
\delta_\omega (M_\pm \times M_\pm) &=& \pm 2(M_\pm^2 + |M_\pm |^2\bI_5)\omega\, .
\end{eqnarray}
The first of these equations implies that\footnote{Notice that  
$\delta_\omega(|M_\pm|^4 - |M_\pm\times M_\pm|^2)=0$, so these particular quartic scalar functions of $(E,B)$ are
Lorentz invariants.} 
\be
 \delta_\omega |M_\pm|^2 = \pm 2\omega \cdot M_- \times M_- \, ,  
\ee
which also follows immediately from \eqref{LBMpm}. 

Observe now that the Lagrangian density of \eqref{6Dus+} can be further rewritten as
\be\label{6Dus++}
\mathcal{L}  =  \frac{\gamma}{2}M_+M_-  + \frac{\lambda}{2} |M_-|^2\, , 
\ee
where 
\be\label{lamdef}
\lambda =  \frac35\lambda^{00} -\upsilon \cdot N\times N +  \tilde\lambda^{ij} (N^2)_{ij} \, , 
\ee
with $N=M_- /|M_-|$, as in \eqref{muom}, and $N^2$ is its matrix square.  This is formally the same as the Lagrangian density \eqref{Lquad} in  the previous subsection, but here $\lambda$ is  not an independent variable, so its Lorentz boost transformation can be calculated from \eqref{LBlms} and \eqref{LBN}, using the 
following formulae analogous to those of  \eqref{formulae}:
\bea
\delta_\omega (N^2)_{ij} &=&  (\omega\cdot N\times N) (2N^2 + \bI_5)_{ij} - \omega_i (N\times N)_j \, ,  \nonumber \\
\delta_\omega (N\times N) &=& 2 (\omega\cdot N\times N) (N\times N) - 2 (N^2 + \bI_5) \omega \, . 
\eea
The result is
\be\label{lamboost-bis}
\delta_\omega\lambda = 2\lambda(\omega \cdot N\times N) \, , 
\ee
which is exactly that of  \eqref{lamboost}, as it had to be because we already know that this is required by Lorentz invariance. 

So far, we have shown that the Lagrangian density \eqref{6DS} of Siegel's 6D chiral 2-form theory can be written in the 
form \eqref{Lquad}, but it still depends on all 20 independent  Lagrange-multiplier components, which (we recall) decompose
into the ${\bf 1}\oplus {\bf 5} \oplus {\bf 14}$ of $SO(5)$.  However, the relation \eqref{lamdef} that  determines the $SO(5)$ scalar $\lambda$ 
in terms of these  ${\bf 1}\oplus {\bf 5} \oplus {\bf 14}$ Lagrange-multiplier components is unchanged if we make arbitrary changes to the 
${\bf 5}\oplus{\bf 14}$ provided we also make an appropriate  change to the ${\bf 1}$, i.e. to $\lambda^{00}$. Specifically, 
$\lambda$ is invariant if 
\be
\frac35\delta_S\lambda^{00} = \left(\delta_S \upsilon\right) \cdot N\times N - \left(\delta_S\tilde\lambda^{ij}\right) (N^2)_{ij} \, , 
\ee
where the subscript $S$ indicates that variations of the Lagrange multipliers satisfying this condition are 
``Siegel symmetry'' gauge transformations in the sense that they are ``trivial''  gauge invariances of the Lagrangian density \eqref{6Dus++}
that allow us to set to zero components of the original Siegel Lagrange multiplier, in this case the  ${\bf 5}\oplus{\bf 14}$ components.  In this gauge we have 
\be\label{Sgf} 
\lambda= \frac35\lambda^{00}\, , 
\ee
and this residual Lagrange multiplier is subject to the gauge transformation \eqref{trivialgt}, which we may now interpret as the residual 
 ``Siegel symmetry''  gauge  transformation that allows us to set $\lambda=0$ almost  everywhere. We have not made any attempt to 
 compare these  ``Siegel symmetry'' gauge  transformations with those given (to first order in an expansion in powers of Lagrange multiplier 
 components)  in \cite{Siegel:1983es}; we presume that that they are equivalent since their effects are equivalent. 

The gauge choice leading to \eqref{Sgf} breaks Lorentz invariance, and this is reflected in the fact that the Lorentz-boost transformation 
of  $\lambda^{00}$ given in \eqref{LBlms} does not agree with the Lorentz 
transformation of ($5/3$ times) $\lambda$ given in \eqref{lamboost-bis}}. However,  this does {\it not} mean that the gauge-fixed Lagrangian 
density is not Lorentz invariant. Instead, it means that a Lorentz transformation must be accompanied by a compensating Siegel 
gauge transformation, with $\omega$-dependent parameters; let us use $\delta_{S(\omega)}$ to denote these
compensating Siegel transformations, which must be such that 
\be
\delta_\omega \upsilon^i + \delta_{S(\omega)}\upsilon^i =0 \, , \qquad 
\delta_\omega \tilde\lambda^{ij} + \delta_{S(\omega)}\tilde\lambda^{ij} =0\, . 
\ee
As $\lambda^{00}$ also transforms under the compensating Siegel transformation, its Lorentz transformation is 
modified to 
\be
\begin{aligned}
\delta'_\omega \left[\lambda^{00}\right] &= \delta_\omega\lambda^{00}  + \delta_{S(\omega)}\lambda^k{}_k \\
&= \delta_\omega\lambda^{00}  + \frac53\left[ \left(\delta_{S(\omega)} \upsilon\right) \cdot N\times N 
- \left(\delta_{S(\omega)}\right)\tilde\lambda^{ij} (N^2)_{ij}\right] \\
&= \delta_\omega\lambda^{00} - \frac53\left[ \delta_\omega \upsilon \cdot N\times N - \delta_\omega \tilde\lambda^{ij} (N^2)_{ij}\right]  \\
&= \frac53 \delta_\omega \lambda\, , 
\end{aligned}
\ee
and hence the relation \eqref{Sgf} is preserved by the combination of a Lorentz transformation and the compensating Siegel 
transformation needed to maintain the gauge choice. 

To summarise, the Lorentz invariance realised linearly on the  ${\bf 1}\oplus {\bf 5} \oplus {\bf 14}$ Lagrange-multiplier components
prior to setting to zero the ${\bf 5} \oplus {\bf 14}$ components (as a partial Siegel gauge choice) becomes a Lorentz invariance
that is realized non-linearly on the surviving singlet Lagrange-multiplier $\lambda^{00}$, and hence on $\lambda$. This partial gauge fixing of Siegel invariances reduces the Lagrangian density \eqref{6Dus+}, 
which is just a rewriting of 6D Siegel Lagrangian density \eqref{6DS},  to the much simpler Lagrangian density of \eqref{Lquad} 
with the single Lagrange  multiplier $\lambda$ and a residual Siegel gauge invariance with transformations given by \eqref{trivialgt}.

\section{Higher dimensions} 

For $2k$-form electrodynamics  in a $(4k+2)$-dimensional locally Minkowski spacetime we have a $(2k+1)$-form field-strength $\mathcal{F}=d\mathcal{A}$ for 
a $2k$-form potential $\mathcal{A}$, which decomposes into $2k$-form potential $A$ and a $(2k-1)$-form potential $\bb{A}$ on the $(2k+1)$-dimensional flat space, 
and we can define analogs of electric and magnetic fields by direct analogy with the $k=1$ case of \eqref{E&B}:
\be\label{E&B2}
\begin{aligned} 
E_{i_1\dots i_{2k}} &:= F_{0i_1\dots i_{2k-1}} \equiv \dot A_{i_1\dots i_{2k}} - 2k \partial_{[i_1} \bb{A}_{i_2 \dots i_{2k}]} \, , \\
B^{i_1\dots i_{2k}} &:= \frac{1}{(2k+1)!}\varepsilon^{i_1\dots i_{2k} j_1\dots j_{2k+1}} F_{j_1\dots j_{2k+1}} \equiv (\nabla \times A)^{i_1\dots i_{2k}} \, . 
\end{aligned}
\ee
Equivalently, 
 \be
 E= \dot A  -(2k) {\rm d}\bb{A} \, , \qquad B = \ast {\rm d}A\, , 
 \ee
where ${\rm d}$ is the exterior derivative on the $(2k + 1)$-space and $\ast$ is the Hodge dual with
respect to its (flat) $(2k+1)$-metric.  The 5-vector algebra used in this paper also generalises: for any 1-form $w$ and 2-forms 
$(C,C^\prime)$, 
 \be
 (w\times C) := \ast (w\wedge C) \, , \qquad C\times C^\prime := \ast(C\wedge C^\prime)\, , 
 \ee
and
\be
C\cdot C^\prime := \frac{1}{(2k)!} C^{i_1\dots i_{2k}} C^\prime_{i_1\dots i_{2k}} \, . 
\ee 
The unique scalar constructible from $(w,C,C^\prime)$ is $\ast(w\wedge C\wedge C^\prime)$, and this can
 be rewritten in either of the two forms given in \eqref{5id}. 
 
 With these definitions/conventions, the HT action for {\it chiral} $2k$-form electrodynamics has a Lagrangian density
that is formally identical to $\mathcal{L}_{HT}$ of \eqref{HTaction}, and its field equation is formally identical to \eqref{6DFJ}. In other 
words, the HT formulation applies for any $k\ge1$ (in addition to reducing to the FJ formulation of the 2D chiral 
boson for $k=0$) \cite{Henneaux:1988gg}. We therefore focus on the Siegel formulation for $k>1$. 

The manifestly Lorentz-invariant Siegel  Lagrangian density for $k>1$ is a straightforward generalization of \eqref{6DS}:
\be
\mathcal {L} = - \frac{\gamma}{2(2k+1)!} \mathcal{F}^{\mu_1\dots\mu_{2k+1}} \mathcal{F}_ {\mu_1\dots\mu_{2k+1}} 
+ \frac{1}{4(2k)!} \lambda^{\mu\nu} \mathcal{F}^+_{\mu\rho_1\dots \rho_{2k}} \mathcal{F}^+_\nu{}^{\rho_1\dots\rho_{2k}} \, , 
\ee
where $\lambda^{\mu\nu}$ is again traceless in the Minkowski spacetime metric. The $k>1$ generalization of \eqref{Fpluscomp} is
\be
\begin{aligned} 
\mathcal{F}^+_{0i_1\dots i_{2k}} &= (E-B)_{i_1\dots i_{2k}} \, , \\
\mathcal{F}^+_{i_1\dots i_{2k+1}} &= - \frac{1}{(2k)!} \varepsilon_{i_1\dots i_{2k+1} j_1\dots j_{2k}} (E-B)^{j_1\dots j_{2k}} \, , 
\end{aligned}
\ee
With respect the $SO(4k+1)$ rotation group, the Siegel Lagrange multiplier decomposes into the following sum of irreps: 
\be\label{reps}
{\bf 1} \oplus {\bf (4k+1)}  \oplus {\bf 2k(4k+3)} \, ,  
\ee
which is the ${\bf 1}\oplus {\bf 5} \oplus {\bf 14}$ decomposition for $k=1$. Using the same notation as we used for the $k=1$ case, we again find that
the Siegel Lagrangian density can be rewriiten in the form \eqref{6Dus++}, except that we now have 
\be
\lambda = \left(\frac{4k-1}{4k+1}\right) \lambda^{00} - \upsilon \cdot  N_- \times N_-  + \tilde\lambda^{ij}\left[N^2\right]_{ij}\, , 
\ee
where $N_- = (E-B)/|E-B|$, as for $k=1$,  but 
\be
\left[N^2\right]_{ij}  = - \frac{1}{(2k-1)!} (E-B)_i{}^{l_1\dots l_{2k-1}} (E-B)_{jl_1\dots l_{2k-1}} \, . 
\ee
For $k=1$ this is the matrix square of $(E-B)$, but the matrix interpretation of $(E-B)$ applies only for $k=1$.  
As for the $k=1$ case, the Siegel gauge invariances may be partially 
fixed by setting to zero all Lagrange multipliers except the singlet $\lambda^{00}$ which is proportional to $\lambda$, and we then have the higher-dimensional version 
of the simple `quadratic' model of subsection \ref{ssec:3quad}. 
 
 Notice that for $k=0$ the formula of \eqref{reps} would give {\it two} Lagrange multipliers rather than one. In light-cone 
 coordinates for the 2D Minkowski spacetime these are the $\lambda^{--}$ and $\lambda^{++}$ components of $\lambda^{\mu\nu}$, but 
 only  the $\lambda^{--}$ component appears in the action (because now $N^2\equiv 1$).

\section{Summary and Discussion}

The original aim of this paper was to revisit Siegel's proposal for manifest Lorentz invariant actions for 
chiral $2k$-form electrodynamics within the context of a Hamiltonian formulation in order to facilitate
comparison with the Henneaux-Teitelboim (HT) formulation, which is first-order and hence essentially 
`already Hamiltonian'. One obvious question of interest is whether the two formulations are equivalent, 
which would allow us to view Siegel's proposal as a means of making manifest the Lorentz invariance 
of the HT formulation. In the $k=0$ case, for which the HT formulation reduces to the Floreanini-Jackiw 
(FJ) formulation of  the 2D chiral boson theory, there appeared to be a consensus  that the Siegel and 
FJ formulations are equivalent.  

However, the claims of equivalence for the Siegel and FJ formulations of the 2D boson are 
at odds with the fact that the FJ chiral boson has an additional gauge invariance since this implies, for periodic 
boundary conditions, that the physical phase spaces are essentially different. This difference has rarely played
a role in discussions of the quantum theory of chiral bosons (e.g. \cite{Sonnenschein:1988ug} which contains a 
useful review of the literature in the immediate aftermath of the work of Floreanini and Jackiw). An exception is the 
much later work of Chen et al. in which the additional gauge invariance of the FJ action is shown to be crucial to 
a  `Lagrangian'  derivation of the partition function for a chiral boson on a 2-torus \cite{Chen:2013gca}. 

Other differences between the 2D Siegel and FJ chiral boson theories are certainly well appreciated. The 
Hamiltonian constraints differ since those of the 2D Siegel theory are first class classically but not quantum 
mechanically, whereas those of the FJ theory are almost all second class. The ``almost'' qualification 
is often omitted but there must be, and there is, one first class constraint to generate the additional 
restricted gauge invariance of the FJ theory \cite{Townsend:2019koy}. It may be that some consistent 
quantization of the 2D Siegel chiral boson will force its quantum equivalence with the FJ chiral boson, 
but any such ``quantization''  is likely to include a procedure that effectively first converts it into the FJ
chiral boson because ``the existence of an additional gauge symmetry is a salient feature of all Lagrangian 
formulations for chiral bosons'' \cite{Chen:2013gca}. In any case, it is our contention that they are 
distinct as {\it classical} theories. 

Most of this paper has been concerned with an examination of these issues for the 6D chiral 2-form 
electrodynamics, for which there are again two distinct  Lagrangian formulations. One is 
Siegel's generalization to 6D of his Lagrangian formulation of the 2D chiral boson, and the other is a
6D generalization by Henneaux and Teitelboim of the FJ chiral boson. These authors have also 
shown that the 6D case is the $k=1$ example of a chiral $2k$-form electrodynamics in a spacetime of 
$(4k+2)$ dimensions,  and that the 2D chiral boson is the $k=0$ case. 

We have pointed out here that the essential physical difference between the Siegel and HT formulations 
is that the HT formulation has additional gauge invariances, first noticed in \cite{Bekaert:1998yp}, and that this 
difference is associated with 
different implementations of the chirality constraint in the Hamiltonian formulation. The linearity of the 
HT formulation means that it is essentially `already'  Hamiltonian; its time-reparametrization  invariant
formulation includes a linear chirality constraint function whose modes are parametrized by time-dependent 
2-forms on the 5-dimensional space. The subset with closed 2-forms is first-class and it generates
a gauge invariance parametrized by closed 2-forms. 

Here we have found a Hamiltonian formulation of Siegel's Lagrangian formulation of chiral 6D 
electrodynamics  in which a simple quadratic chirality constraint is a imposed by a rotation-scalar Lagrange 
multiplier $\lambda$ that is a non-linear function of the phase-space fields and the Lagrangian Lagrange multipliers.
The ``trivial'' Siegel-symmetry gauge invariances of the Lagrangian formulation can now be used to set to zero all but one 
component, proportional to $\lambda$ in this gauge, on which Lorentz invariance is now non-linearly realized. 
The remaining Siegel-symmetry gauge invariance is generated by the chirality constraint function, while the 
2-form gauge transformation is generated as for the non-chiral theory, with a parameter that is an 
exact 2-form.

We have also shown how this difference generalises to chiral $2k$-form electrodynamics. In the Hamiltonian 
formulation, the ``additional'' gauge invariances of the HT  action are parametrized by $2k$ forms on the 
$(4k+1)$-dimensional space that are closed but not exact. These
are elements of the de Rham cohomology group  $H^{2k}(X)$, where $X$ is the $(4k+1)$-dimensional 
space. This result applies equally for $k=0$, for which $X$ is either $\mathbb{R}$ or $S^1$,  for which $H^0(X)$ 
has dimension $0$ and $1$ respectively; the one additional gauge invariance for $X=S^1$ is the additional 
gauge invariance of the FJ chiral boson theory with periodic boundary conditions. 

If $H^{2k}(X)$ is trivial, there is no physical difference between the field equation $E=B$ and the HT 
field equation $\dot B =\nabla\times B$, as pointed out in \cite{Bekaert:1998yp}. This is consistent with the fact, 
established in \cite{Buchbinder:2020ocz},  that the unique unitary irreducible representation of the 6D Poincar\'e group 
specified by zero mass and spin in the $({\bf 3},{\bf 0})$ irrep of the $SO(4)$ ``little group''  corresponds to the free-field theory 
with field equation $E=B$ because the 6D Lorentz subgroup is broken by identifications whenever $H^{2}(X)$ is non-trivial (and an analogous 
observation applies for any $k>1$). Such identifications preserve the local metric structure of spacetime, which is 
therefore locally (but not globally) Minkowski; it is the  ``local'' qualification that  has allowed us to consider periodic 
boundary conditions (and hence closed strings for 2D and toroidal compactifications for higher dimensions). 

We have restricted our analysis in this paper to free-field chiral $2k$-form electrodynamics in a 
$(4k+2)$-dimensional  locally-Minkowski spacetime. These restrictions have been imposed 
for reasons of simplicity of presentation; we expect a generic classical inequivalence of the Siegel and HT formulations 
to survive the introduction of self-interactions (e.g, in the context of M5-brane dynamics for $k=1$) or gravitational 
interactions leading to spacetimes that are not locally Minkowski.


\section*{Acknowledgements}
This work has been partially supported by STFC consolidated grant ST/T000694/1.

\providecommand{\href}[2]{#2}\begingroup\raggedright\endgroup


\end{document}